\documentclass[10pt, letter, a4paper]{./lib/IEEEtran}
\usepackage[cmex10]{amsmath}
\usepackage{mathtools}
\usepackage{array}
\usepackage{algpseudocode,algorithm,algorithmicx}
\usepackage{float}
\usepackage{tabularx}
\usepackage{mathrsfs} 
\usepackage{tikz}
\usetikzlibrary{chains, arrows,shapes,positioning,arrows,decorations.markings,calc}
\usepackage{pgfplots}
\pgfplotsset{compat=newest}
\usepgfplotslibrary{groupplots}
\usepgfplotslibrary{fillbetween}
\usepackage{gensymb}
\usepackage{xcolor}
\usepackage[shortcuts,acronym]{glossaries}
\makeglossaries
\usepackage{fixmath}
\usepackage{amssymb}
\usepackage{bm}
\usepackage{bbm}
\usepackage{pbox}
\renewcommand{\arraystretch}{2}	
\usepackage{tabularx}
\pgfplotsset{compat=newest,
	/pgfplots/ybar legend/.style={
		/pgfplots/legend image code/.code={%
			\draw[##1,/tikz/.cd,bar width=3pt,yshift=-0.2em,bar shift=0pt]
			plot coordinates {(0cm,0.8em)};},
	},}

\newcommand\myeqa{\stackrel{\mathclap{\mbox{($a$)}}}{=}}
\newcommand\myeqb{\stackrel{\mathclap{\mbox{($b$)}}}{=}}
\DeclarePairedDelimiter{\ceil}{\lceil}{\rceil}

\usepackage{amsthm}
\theoremstyle{plain}
\newtheorem*{theorem*}{$(\bm{\theta }, \bm{\alpha})$-\textit{availability}}
\usepackage{caption}

\begin{document}
%
\title{Availability and Reliability of Wireless Links in 5G Systems: A Space-Time Approach}
\author{\IEEEauthorblockN{Mustafa Emara, Miltiades C. Filippou, Ingolf Karls\\}
	\IEEEauthorblockA{Next Generation and Standards, Intel Deutschland GmbH, Neubiberg, Germany \\
		Email:$\{$mustafa.emara, miltiadis.filippou, ingolf.karls$\}$@intel.com}
}
\maketitle
\thispagestyle{empty}
\maketitle
\thispagestyle{empty}

\newacronym{AP}{AP}{access point}
\newacronym{APs}{APs}{access points}
\newacronym{CTMC}{CTMC}{continous time Markov chain}

\newacronym{EPs}{EPs}{end points}
\newacronym{5G}{5G}{fifth generation}
\newacronym{IoT}{IoT}{internet of things}
\newacronym{KPIs}{KPIs}{key performance indicators}
\newacronym{QoS}{QoS}{quality of service}
\newacronym{RAT}{RAT}{radio access technology}
\newacronym{SIR}{SIR}{signal to interference ratio}
\newacronym{URLLC}{URLLC}{ultra reliable low latency communication}
\thispagestyle{empty}
\begin{abstract}
Wireless links are characterized by fluctuating quality leading to variable packet error rates which are orders of magnitude higher than the ones of wired links. Therefore, it is of paramount importance to investigate the limitations of using 5G wireless links for internet of things (IoT) applications. 5G wireless links in IoT need to assure determinism of process flows via real-time communication \emph{anytime} and \emph{anywhere}, which is an utmost requirement for multiple verticals like automotive, industrial automation and aerospace. Based on a space-time approach, in this work, we provide novel definitions of wireless link availability and reliability, assuming a number of access points (APs) and end points (EPs) deployed over a fixed area. Our objective is to analyze the availability of a service in both domains. In the space domain, we characterize spatially available areas consisting of all locations that meet a performance requirement with confidence. In the time domain, we propose a channel allocation scheme accounting for the spatial availability of a cell. To emphasize the incurred space-time performance trade-offs, numerical results are presented, also highlighting the effect of different system parameters on the achievable link availability and reliability. 
\end{abstract}
%

\section{Introduction}\label{section:introduction}
Due to the increasing service requirements posed by verticals ready to exploit \ac{5G} wireless communication systems, enhancing existent \ac{KPIs} and defining new ones is inevitable \cite{3GPP2018}. Two important performance requirements are the reliability and availability of communication, which need to be satisfied in an end-to-end manner \cite{NGMNA2016}. Among the established \ac{5G} service categories,  \ac{URLLC}  shall pave the way for \ac{IoT}, mission critical services, real-time control and automation for multiple market segments \cite{Popovski2014}. However, \ac{URLLC} is accompanied by a plethora of challenges and stringent requirements to ensure an ongoing service with virtually no failures during the operation time. These challenges are currently being addressed by academia \cite{Bennis2018} and different industry associations/ standardization bodies \cite{3GPP2018}, \cite{2018}, \cite{2017}. For instance, \ac{URLLC} is envisaged to support a packet failure probability of $10^{-7}$, while introducing latencies of merely up to the few milliseconds range \cite{Popovski2014}.  Since such performance demands are challenging for legacy wireless networks, a paradigm shift from the conventional network architecture is inevitable. 

In both recent 3rd Generation Partnership Project (3GPP) specifications and academic research works, \ac{URLLC} service performance has been mainly evaluated via investigating metrics such as packet error ratio, latency and jitter \cite{3GPP2018}, \cite{Popovski2014}. These metrics, though fundamentally meaningful from the radio communication perspective, need to be looked at together with service demands from a vertical-specific point of view (e.g., availability of a service and reliability of the operation). Consequently, such service-specific metrics need to be first well defined, understood and then mapped to the wireless system's parameters, prior to evaluating system-wide feasibility of the focused service/operation. Conceptually, this novel system view aims to unlock the potential of running wireless services quasi-deterministically, thus, enabling a fine-grained system analysis \cite{Bennis2018}. 

\subsection{Related Work}
To the best of our knowledge, adopting definitions of service-tailored link availability and reliability for wireless-based systems has not yet been expressed adequately. In \cite{Mendis2017}, the authors have proposed a new definition of spatial availability, as the ratio of the mean covered area to the geographical area of a given \ac{AP}. Nevertheless, an interference-free scenario was considered and no insights on the time evolution of communication availability were provided. Additionally, in \cite{Balapuwaduge2018}, the authors proposed a reliability metric, consisting of two components: the temporal availability and the probability to overcome a received power threshold, however, for a single cell scenario. Furthermore, the authors in \cite{Hoessler2017} summarized main definitions from reliability theory \cite{Birolini2010}, and presented an automation-based use case exploiting multi-link connectivity. Nevertheless, spatial availability analysis was not considered at all. In addition, the authors in \cite{Bennis2018}, provided an tutorial-like overview, introducing different challenges and solution proposals for \ac{URLLC} services. Although quite insightful, this work did not touch upon the concepts of time and space availability. Finally, \cite{Shrestha2018} proposed a new protocol enabling precise synchronization for wireless time sensitive networking. Although no definitions of reliability or availability were discussed in this work, the proposed framework aims to achieve operation determinism in wireless systems.  
\subsection{Contributions}
Motivated by the above, in this paper, concentrating on both space and time domains, we make a first attempt to bridge the gap between traditional radio link \ac{KPIs} and service-level \ac{KPIs} by providing an insight on the availability and reliability of wireless links in \ac{5G} systems. The proposed framework aims to indicate which locations in a given area would overcome a performance threshold over a specific time window with a guaranteed level of confidence. In further detail, the contributions of the paper are the following
\begin{itemize}
	\item Focusing on a \ac{RAT}-agnostic wireless system, we propose the definition of a new, stochastic quantity to measure the spatial availability of a wireless link for a service-relevant confidence level.
	\item Capitalizing on the proposed definition of spatial availability, we propose a novel resource allocation approach dependent on the spatially available area of a given \Ac{AP} and based on the concept of resource provisioning. In addition, transient and steady-state analysis of the system's temporal availability is performed and the relation between temporal availability and reliability is highlighted.
	\item We present numerical evaluations, highlighting the different effects of system parameter values on spatial availability, as well as on temporal availability and reliability. In addition, we show the relation between spatial and steady state temporal availability.
\end{itemize}
\subsection{Notation}
Throughout this paper, we adopt the following notation. Matrices and vectors are represented as upper-case and lower-case boldface letters ($\bm{A}$, $\bm{a}$), respectively. Also, $f_{\bm{i}}$ denotes the value of quantity $f$ at location $\bm{i}$ of Cartesian coordinates $(x, y)$. The operation $\ceil[\big]{a}$ represents the rounding of $a$, $\mathbbm{1}(a)$ is the indicator function, which equals 1 if $a$ is true and 0 otherwise and min($a,b$) is the minimum operator of $a$ and $b$. Moreover,  $\mathbb{P}[A]$ is the probability of event $A$, $\mathbb{E}[X]$ is the expectation of random variable $X$, $M_X(.)$ is the moment generating function (MGF) of $X$, and finally, the operation $d(\bm{x},\bm{y})$ represents the Euclidean norm between $\bm{x}$ and $\bm{y}$. 

\thispagestyle{empty}
\section{System Model}\label{sec:system_model}
\begin{figure}
	\begin{center}
		\includegraphics[width=\columnwidth]{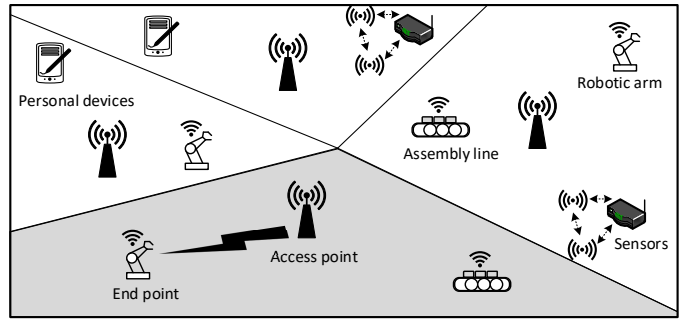}
		\caption{The investigated system model consisting of different \ac{EPs} types, each served by a single AP. The shaded area represents the connectivity region of an AP.}
		\label{fig:system_model}
	\end{center}
\end{figure}

In this paper, a downlink wireless system, consisting of $N$ single-antenna \ac*{APs} of equal transmit power is deployed over a two-dimensional bounding box (e.g., a factory floor). It is assumed that each spatial deployment realization of the $N$ \ac{APs} is denoted as $\Phi$ (i.e., $\{l|\;(x_l,y_l) \in \mathbb{R}^2,\;l\in \Phi  \}$). Without loss of generality, the proposed system model can be applied to different communication systems. A multitude of \ac{EPs}, like personal tablets, control units, sensors or actuators,  are being served via wireless links, as shown in Fig. \ref{fig:system_model}. Each \ac{AP} has access to $M$ orthogonal channels that can be used for downlink transmission, so as for the EPs to fulfill their service requests. At the EP side, service requests form an arrival process which follows a Poisson distribution with an average arrival rate denoted by $\lambda$, whereas the service time of a downlink transmission follows an exponential distribution with an average service rate of $\mu$. A frequency reuse factor of one is assumed in this work, which translates to the potential presence of inter-cell interference among the $N$ \ac{APs}. From a joint deployment and connectivity point of view, the cell's connectivity region (i.e., the shaded area in Fig. \ref{fig:system_model}, also named as  Voronoi cell) represents the geographical area in which a wireless link can be established between an EP and its closest \ac{AP}. Equivalently, Voronoi cells are shaped by applying an EP-AP connectivity rule based on a minimum pathloss criterion.

An important metric to evaluate \ac{QoS} in interference-limited scenarios, is the \ac{SIR} \cite{Andrews2011}. The \ac{SIR} of a generic EP located at point $\bm{i}$ and served by an \ac{AP} located at point $\bm{j}$, where $\bm{i}$ and $\bm{j}  \in \mathbb{R}^2$, is computed as   
\begin{equation}
\text{SIR}_{\bm{i}, \bm{j}} = \frac{P_{\text{Tx}} h_{\bm{i}, \bm{j}} \mathcal{L}(\bm{i}, \bm{j}, \eta)}{\sum_{\bm{k} \in \mathcal{I}}^{} P_{\text{Tx}} h_{\bm{i}, \bm{k}}\mathcal{L}(\bm{i}, \bm{k}, \eta)},
\end{equation}
where $P_{\text{Tx}}$ is the transmit power of the \ac{AP}, $h_{\bm{i},\bm{j}}$, $h_{\bm{i},\bm{k}}, k \in \mathcal {I}$ are the small-scale fading parameters, $\eta$ is the pathloss exponent and $\mathcal{I}$ represents the set of all the $N-1$ interfering \ac{APs}. The function $\mathcal{L}(.)$ computes the pathloss attenuation between two points $\bm{i}, \bm{j} \in \mathbb{R}^2$ as follows: $\mathcal{L}(\bm{i}, \bm{j}, \eta) = (d(\bm{i}, \bm{j}))^{-\eta}$. It should be noted that each \ac{AP} is aware of the locations of interfering \ac{APs}. Additionally, the fast fading parameters are assumed to be Rayleigh distributed with unit average power, i.e., for every \ac{AP}-EP link, $h\approx\text{exp}(1)$ and the fast fading effects are assumed non-correlated among the various links. 
To investigate the \textit{guaranteed performance} of a given \ac{AP}-EP link, we introduce a new binary evaluation metric, $\Omega_{\bm{i}, \bm{j}}(\theta, \alpha, \Phi)$, having as a decision criterion the probability for a wireless link to achieve a given \ac{SIR} threshold $\theta$ with a predetermined confidence level $\alpha$ for a given \ac{AP} deployment realization $\Phi$ \cite{Haenggi2016}. This quantity is mathematically expressed as follows
\begin{equation}\label{eq:fine_grainedSIR}
\Omega_{\bm{i}, \bm{j}}(\theta, \alpha,\Phi) = \mathbbm{1}(\mathbb{P}[\text{SIR}_{\bm{i}, \bm{j}} \ge \theta] \ge \alpha).
\end{equation}
This metric will be exploited in the coming section for defining the spatial, service-relevant availability of a wireless link.
\thispagestyle{empty}
\section{Availability Analysis: Spatial Domain}\label{sec:spatialanalysis}
Throughout this section, we aim to project the well-established definitions of time-domain availability and reliability to the spatial domain. Temporally, instantaneous availability of a system is the probability of the system being operational at a given time instant \cite{Birolini2010}, whereas, in the space domain, as introduced in \cite{Mendis2017}, the spatial availability $A_s$, defines the locations on a given Euclidean plane, where the system is operational. The region of operation was modeled as the cell's circular coverage area in \cite{Mendis2017}, due to the lack of interference, whereas, in \cite{Mendis2017a}, the authors proposed multiple criteria for the definition of spatial availability. Inspired by these two works, in this section we propose a new, service-related definition of spatial availability, taking into account the confidence level of surpassing a predefined \ac{SIR} threshold. We present the following definition:
\begin{theorem*}
	Any EP located at $\bm{i}$ and served by an \ac{AP} located at $\bm{j}$ is labeled as $(\theta, \alpha)$-\textit{available}, if $\Omega_{\bm{i}, \bm{j}}(\theta, \alpha, \Phi) = 1,\; \bm{j} \in \Phi,\;\bm{i},\bm{j}\in \mathbb{R}^2,$ and \textit{non-available} otherwise.
\end{theorem*}
Given an AP deployment $\Phi$ and accumulating all \ac{EPs} possible locations $\bm{z}, \bm{z} \in \mathbb{R}^2$ which satisfy the $(\theta, \alpha)$ spatial availability criterion $\Omega_{\bm{z},\bm{j}}(\theta, \alpha, \Phi)$ when connected to an AP located at point $\bm{j}$, we obtain the following $(\theta, \alpha)$-available region $\mathcal{D}_j $ as follows
\begin{equation}\label{eq:set_D}
\mathcal{D}_j  = \{ \bm{z} \in \mathbb{R}^2 | \bm{z} \in \mathcal{D}_{\bm{j}},\; \Omega_{\bm{z}, \bm{j}}(\theta, \alpha, \Phi) = 1,\; \bm{j} \in \Phi\}.
\end{equation}
Thus, the proposed spatial availability for an \ac{AP} located at $\bm{j}$, can be formulated as  
\begin{align}\label{eq:A_s}
A_s(\bm{j}) &= \text{min}\Big(1,\frac{\text{Area}(\mathcal{D}_{\bm{j}})}{\text{Area}(\mathcal{V}_{\bm{j}})}\Big) = \text{min}\Big(1, \frac{|\mathcal{D}_{\bm{j}}|}{|\mathcal{V}_{\bm{j}}|}\Big),
\end{align}
where the minimum operator accounts for cases where the ($\theta, \alpha$)-available area is larger than geographical area of the \ac{AP} (in such cases $A_{s}(\bm{j})$ = 1) and $\mathcal{V}_{\bm{j}}$ is the collection of points constituting the Voronoi cell of the $j$-th \ac{AP}. In addition, eq. (\ref{eq:A_s}) can be expanded as
\begin{equation}
\text{min}\Big(1,\frac{\int_{\bm{z}\in\mathbb{R}^2}\mathbbm{1}(\bm{z}\in\mathcal{D}_{\bm{j}}) \text{d}\bm{z}}{\frac{1}{2}|\sum_{l}^{g-1}x_ly_{l+1} + x_qy_q -\sum_{l}^{g-1}x_{l+1}y_l - x_qy_q|}\Big),
\end{equation}
where $g$ is the number of edges and ($x_l,y_l$) are the Cartesian coordinates of the $l$-th vertex of the Voronoi cell. The denominator in eq. (\ref{eq:A_s}) is obtained by applying the well-known shoelace algorithm that computes the area of a Voronoi polygon with $g$ edges \cite{Mendis2017}. 

In order to compute the area of set $\mathcal{D}_j$, its boundary needs to be specified. In other words, focusing on an \ac{AP}, all the points satisfying the \ac{SIR} threshold $\theta$ with confidence level $\alpha$ are sought. One can thus expand eq. (\ref{eq:fine_grainedSIR}) \cite{Ganti2010}, for a given spatial deployment ($\Phi$ is dropped for simplicity) as follows
\begin{align}\label{eq:spatial_avail}
	\Omega_{\bm{z}, \bm{j}}(\theta, \alpha) &= \mathbbm{1}\Big(\mathbb{P}\Big[\frac{P_{\text{Tx}} h_{\bm{z}, \bm{j}} \mathcal{L}(\bm{z}, \bm{j}, \eta)}{\sum_{\bm{k} \in \mathcal{I}}^{} P_{\text{Tx}} h_{\bm{z}, \bm{k}}\mathcal{L}(\bm{z}, \bm{k}, \eta)} \ge \theta \Big] \ge \alpha\Big), \nonumber \\
	&\myeqa  \mathbbm{1}\Big(\mathbb{E} \Big[ \text{exp} \Big(\frac{-\theta}{\mathcal{L}(\bm{z}, \bm{j}, \eta)} \sum_{\bm{k} \in \mathcal{I}}^{} \mathcal{L}(\bm{z}, \bm{k}, \eta \Big) \Big] \ge \alpha\Big), \nonumber \\
	&\myeqb \mathbbm{1}\Big(\prod_{k \in \mathcal{I}} M_h \Big(\frac{-\theta }{\mathcal{L}(\bm{z}, \bm{j}, \eta)}\mathcal{L}(\bm{z}, \bm{k}, \eta) \Big) \ge \alpha\Big),\nonumber \\
	&\myeqa \mathbbm{1}\Big(\prod_{k \in \mathcal{I}} \frac{1}{1+\frac{\theta }{\mathcal{L}(\bm{z}, \bm{j}, \eta)}\mathcal{L}(\bm{z}, \bm{k}, \eta)} \ge \alpha\Big),
\end{align}
where $(a)$ follows since fast-fading channels are assumed to be exponentially distributed and $(b)$ is the MGF of an exponentially distributed random variable. Since the \ac{APs} transmit with equal power, the final expression of $\Omega_{\bm{z}, \bm{j}}(\theta, \alpha)$ is transmit power independent \cite{Andrews2011}. 

Expression eq. (\ref{eq:spatial_avail}) can be utilized to define the boundary of Area($\mathcal{D}_j$) by substituting inequality by pure equality. However, since this boundary is hardly tractable in closed form, to obtain quantitative results, we resort to a bisection-based algorithm to approximate the size of this area to be then used in eq. (\ref{eq:A_s}).
\begin{figure} 
	\centering
	\input{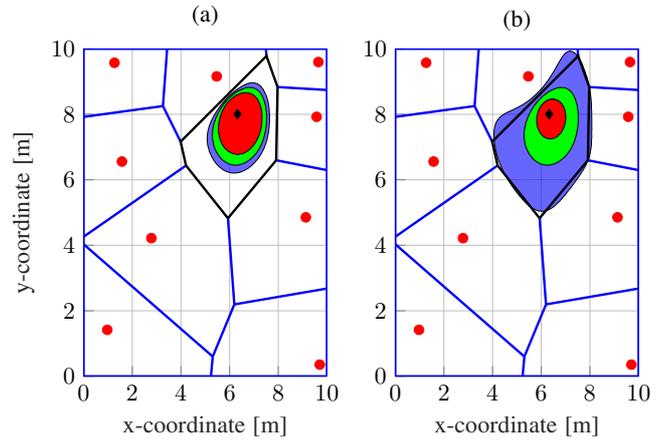}
	\vspace{-7pt}
	\caption{$(\theta, \alpha, \Phi)$-available regions for a deployment with $N=10$ APs focusing on a generic \ac{AP} (its Voronoi border drawn in black) (a) $\theta = 0$ dB and $\alpha=( \textcolor{blue}{0.7}, \textcolor{green}{0.8}, \textcolor{red}{0.9})$ and (b) $\theta = ( \textcolor{blue}{-10}, \textcolor{green}{0}, \textcolor{red}{10})$ dB and $\alpha = 0.8$.}
	\label{fig:coverge_regions} 
\end{figure}
A visualization of the computed regions is shown in Fig. \ref{fig:coverge_regions}, where different combinations of $(\theta, \alpha)$ are considered for a given \ac{AP}.
It is observable that in Fig. \ref{fig:coverge_regions}(b), the $(-10\text{ dB}, 0.8)$ region is not convex, due to the interference imposed by the closest interfering \ac{AP}, which reveals that, along with $(\theta, \alpha)$, the number and location of deployed and, consequently, interfering \ac{APs} is expected to highly affect the spatial availability. Thorough investigation on the effect of $(\theta, \alpha)$ on spatial availability for random \ac{AP} deployments will be considered in Section \ref{sec:simulation_results}.

\thispagestyle{empty}
\section{Availability Analysis: Temporal Domain}\label{sec:temporalanalysis}
Having analyzed the spatial availability metric in the previous section, and since our objective is to propose a unified, space-time availability framework useful to \ac{URLLC} systems, in this section we concentrate on the time domain.
\subsection{Resource Partitioning}
As explained in Section \ref{sec:system_model}, each \ac{AP} has $M$ channels that can be accessible by the \ac{EPs} in its Voronoi region. To account for the spatial availability $A_s$ as defined in Section \ref{sec:spatialanalysis}, we propose a spatial availability-proportional channel allocation scheme. According to this scheme, since $A_s$ decomposes the Voronoi region of the AP into two regions, the number of channels to be utilized by \ac{EPs} located in the $(\theta, \alpha)$-available and non-available regions of an \ac{AP} located at point $\bm{j}$ can be, respectively, written as  
\begin{equation}\label{eq:channel_access}
M_a(\bm{j}) = \ceil[\big]{A_s(\bm{j})M},\; M_n(\bm{j}) = M - M_a(\bm{j}).
\end{equation}
As a result of the proposed policy, assuming that service requests arrive uniformly in space, when the spatial availability ratio $A_s$ is low, a few channels will be allocated to the few evolving requests coming from the $(\theta, \alpha)$-available region, while, the majority of channels will be allocated to the (possibly many) requests coming from the $(\theta, \alpha)$ non-available region. 
\begin{figure}
	\begin{center}
		\input{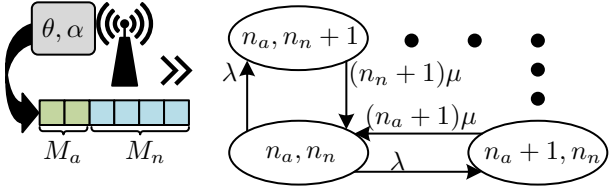}
		\caption{Resources partitioning based on $A_s$ along with part of the two dimensional birth/death Markov process.}
		\label{fig:channelaccess}
	\end{center}
\end{figure}
\subsection{Temporal Analysis}
In order to model the time-dependent status of the resources at a generic \ac{AP}, we resort to a \ac{CTMC} model that captures the number of idle/busy channels as time evolves. To also capture the decomposition of service requests into two sets (spatially available/non-available), a two dimensional \ac{CTMC} is employed, where the first dimension represents the number of \ac{EPs} being served within the $(\theta, \alpha)$-available region of the \ac{AP}, and the other dimension represents the number of \ac{EPs} in the rest of the geographical cell region. Fig. \ref{fig:channelaccess} presents the proposed framework, where $n_a$ and $n_n$ are generic numbers of \ac{EPs} being served in the two mentioned regions. Such a model leads to a finite birth/death Markov process, where the total number of states is limited by the total number of channels and all possible partitioning options. Thus, for a given channel allocation, the set of feasible states can be represented as 
\begin{align}
\mathcal{S} = &\{(n_a, n_n)| \; 0 \le n_a \le M_a,\; 0 \le n_n \le M_n,\; \nonumber\\ 
&M_a + M_n = M \},
\end{align}
where the total number of states is $|\mathcal{S}| = (M_a+1)(M_n+1)$, since a number of $n$ channels will lead to $n+1$ states. 
Based on the above described model,  the \textit{temporal availability} is defined as the probability of at least one channel being available for a new request. As a result, the set of temporally available states for the  $(\theta, \alpha)$-available and non-available regions are $\mathcal{A}^a = \{(n_a, n_n), n_a = \{ 0,1,\cdots,M_a -1\}\}$ and $\mathcal{A}^n = \{(n_a, n_n), n_n = \{ 0,1,\cdots,M_n -1\}\}$, respectively. The state equations can be vectorized as $\bm{\tau}(t) =\{ \tau_1(t), \tau_2(t), \cdots, \tau_{|\mathcal{S}|}(t) \}$, where $\tau_l(t)$ is the probability of the system being in  the $l$-th state at time instant $t$. Resorting to the matrix notation and using the Kolmogorov forward equations \cite{Birolini2010}, the state probabilities can be computed by solving the following equation 
\begin{equation}\label{eq:state_probability}
\frac{\text{d}}{\text{d}t}\bm{\tau}(t) = \bm{\tau}(t)\bm{Q},
\end{equation}
where the infinitesimal generator (i.e., transition rate) matrix is denoted by $\bm{Q}$, with dimension $|\mathcal{S}|\times|\mathcal{S}|$. To compute the system's temporal availability, one needs to solve eq. (\ref{eq:state_probability}). We adopted a similar approach as in \cite{Balapuwaduge2017}, based on the uniformization method \cite{Boucherie1998}, where the solution of eq. (\ref{eq:state_probability}), for a given initial state probability (i.e., $t=0$), denoted by $\bm{\tau}(0)$, can be rewritten as 
\begin{align}\label{eq:state_probability2}
\bm{\tau}(t) &= \bm{\tau}(0)\text{e}^{\bm{Q}t} = \bm{\tau}(0)\sum_{i=0}^{\infty} \frac{(\bm{Q}t)^i}{i!}, \nonumber \\
&\myeqa \bm{\tau}(0)\text{e}^{-qt} \sum_{i=0}^{\infty} \frac{(qt)^i}{i!}\bm{R}^n, 
\end{align}
where  $(a)$ follows from the introduction of $\bm{R}=\bm{I} + \frac{1}{q}\bm{Q}$, $\bm{I}$ is the identity matrix and $q$ is a number satisfying $q\ge\text{max}(q_{ii})$, where $q_{ii}$ are the diagonal elements of $\bm{Q}$. To numerically solve eq. (\ref{eq:state_probability2}), the summation must be truncated at level $N_c$ as shown in \cite{Balapuwaduge2017}. In order to obtain the system's temporal availability at a given time instant $t$, one needs to consider all the available states as follows
\begin{equation}\label{eq:temp_avail}
A_t^u(t) = \sum_{i\in\mathcal{A}^u} \bm{\tau}_i(t),\; u\in\{a,n\}, 
\end{equation}
where index $u\in\{a,n\}$ represents the $(\theta, \alpha)$-available and non-available regions, respectively.
\subsubsection{Reliability Analysis}
 Another important metric for the temporal analysis is the system's \textit{temporal reliability} $R(t)$ \cite{Hoessler2017}, which is defined as the probability that the system is operational during time interval $[0,t]$. Such a definition can be employed  in the studied \ac{CTMC} model, by forcing the system to remain in an unavailable state once it reaches one. In other words, the transition rate from unavailable state is set to zero \cite{Hoessler2017}. Such a modification leads to a modified infinitesimal generator matrix $\hat{\bm{Q}}$ and eq. (\ref{eq:state_probability}) can be re-expressed as 
\begin{equation}
\frac{\text{d}}{\text{d}t}\hat{\bm{\tau}}(t) = \hat{\bm{\tau}}(t)\hat{\bm{Q}},
\end{equation} 
where $\hat{\bm{\tau}}(t)$ corresponds to state probability of the modified \ac{CTMC}. Accordingly, the system's temporal reliability is computed as
\begin{equation}
R^u(t) = \sum_{i\in\mathcal{A}^u} \hat{\bm{\tau}}_i(t),\; u\in\{a,n\}.
\end{equation}
As it will be numerically shown later, the system reliability is always upper bounded by its time availability (i.e., $A_t(t)\ge R(t)$), since for a repairable system, transition rates from a failed state are non-zero.
\subsubsection{Steady State Analysis}
Another interesting metric relevant to temporal analysis is the steady state time availability, which is time independent and can be interpreted as the average operating time \cite{Birolini2010}. Mathematically \cite{Hoessler2017}, it can be represented as 
\begin{align}
A_t^u = \lim_{t\to\infty} A_t^u(t) = \sum_{i\in\mathcal{A}^u} \bm{\tau}_i 
= \sum_{i\in\mathcal{A}^u} \frac{\rho_i}{i!}\Big(1+ \sum_{l=1}^{M_u} \frac{\rho^l}{l!}, \Big),
\end{align}
where  $u\in\{a,n\}$ and $\rho = \frac{\lambda}{\mu}$ represents the arrival to service rate ratio. 
\thispagestyle{empty}
\section{Simulation Results}\label{sec:simulation_results}
The objective of this section is to highlight how different wireless system parameters affect the system's space-time availability and reliability, as defined in Sections \ref{sec:spatialanalysis} and \ref{sec:temporalanalysis}. The values of the involved system parameters are provided in Table \ref{Table:simulation_parameters}, unless stated otherwise. 
\begin{table}
	\centering
	\caption{Simulation Parameters.}
	\renewcommand{\arraystretch}{1}
	\resizebox{0.9\columnwidth}{!}{
		\begin{tabular}{| l | l |}
			\hline
			\textbf{Parameter} & \textbf{value}  \\ \hline \hline
			Deployment area & 100 m$^2$ \\
			Pathloss exponent ($\eta$) & 4 \\
			Total channels ($M$) & 10 \\
			Initial state probability ($\bm{\tau}(0)$) & $\bm{0}$ \\
			Requests arrival rate ($\lambda$) & 8 packets/sec\\
			Requests service rate ($\mu$) & 1 packet/sec \\
			Spatial realizations & 10000 \\ \hline
	\end{tabular}}
	\label{Table:simulation_parameters}
\end{table}
\subsection{Spatial Analysis}
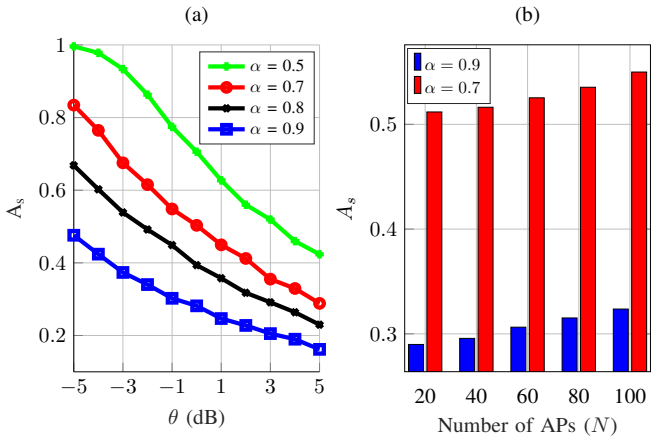
\begin{figure} 
	\centering
	\captionsetup{width=\columnwidth}
	\definecolor{mycolor1}{rgb}{0.00000,0.44700,0.74100}%
\definecolor{mycolor2}{rgb}{0.85000,0.32500,0.09800}%
\definecolor{mycolor3}{rgb}{0.92900,0.69400,0.12500}%

\begin{tikzpicture}[scale=0.8]
\begin{groupplot}[group style={
	group name=myplot,
	group size= 2 by 1, horizontal sep=1.4cm},height=3cm, width= 0.85*\columnwidth]
\nextgroupplot[title={(a)}, align=left,
width  = 0.31*\textwidth,
height = 7cm,
xmin=-5,
xmax=5,
xlabel style={font=\color{white!15!black}},
xlabel={$\theta$ (dB)},
ymin=0.1,
ymax=1,
ylabel style={font=\color{white!15!black}},
ylabel={$\text{A}_\text{s}$},
axis background/.style={fill=white},
axis x line*=bottom,
axis y line*=left,
xtick={-5, -3,  -1, 1, 3, 5},
xmajorgrids,
ymajorgrids,
legend style={nodes={scale=0.8}, at={(0.98,0.98)},anchor=north east, draw=white!15!black}
]
\addplot [color=green, line width=2.0pt, mark=+, mark options={solid, green}]
table[row sep=crcr]{%
	-5	0.995861536741932\\
	-4	0.978248263123069\\
	-3	0.933445148115368\\
	-2	0.863077476942397\\
	-1	0.774401852485676\\
	0	0.705812479040648\\
	1	0.627153052487516\\
	2	0.560065381043838\\
	3	0.519284284801083\\
	4	0.459756936174509\\
	5	0.423484678029887\\
};
\addlegendentry{$\alpha\text{ = 0.5}$}

\addplot [color=red, line width=2.0pt, mark=o, mark options={solid, red}]
table[row sep=crcr]{%
	-5	0.834307382701798\\
	-4	0.764799977951048\\
	-3	0.6754196360202\\
	-2	0.615333697006347\\
	-1	0.548307883623795\\
	0	0.50329777917356\\
	1	0.4495588794183\\
	2	0.412073840869267\\
	3	0.35522209587813\\
	4	0.329424714929261\\
	5	0.288396819187204\\
};
\addlegendentry{$\alpha\text{ = 0.7}$}

\addplot [color=black, line width=2.0pt, mark=x, mark options={solid, black}]
table[row sep=crcr]{%
	-5	0.66835210994693\\
	-4	0.602512536556543\\
	-3	0.538780448000238\\
	-2	0.491895571296631\\
	-1	0.448881845789908\\
	0	0.393723469652905\\
	1	0.357649997637481\\
	2	0.317620667247957\\
	3	0.291371542541644\\
	4	0.263911303419706\\
	5	0.22976166170086\\
};
\addlegendentry{$\alpha\text{ = 0.8}$}

\addplot [color=blue, line width=2.0pt, mark=square, mark options={solid, blue}]
table[row sep=crcr]{%
	-5	0.476001624251499\\
	-4	0.423949159424402\\
	-3	0.373611466554888\\
	-2	0.34023989317506\\
	-1	0.302413309935674\\
	0	0.281639621962343\\
	1	0.246691471007822\\
	2	0.227703710782475\\
	3	0.205215643538586\\
	4	0.189760581567721\\
	5	0.161816294978015\\
};
\addlegendentry{$\alpha\text{ = 0.9}$}

\nextgroupplot[title={{(b)}},
width  = 0.31*\textwidth,
height = 7cm,
major x tick style = transparent,
xlabel style={font=\color{white!15!black}},
xlabel={Number of APs ($N$)},
ylabel={$A_s$},
ybar=4*\pgflinewidth,
bar width=7pt,
ymajorgrids = true,
xmajorgrids = true,
symbolic x coords={20, 40, 60, 80, 100},
scaled y ticks = false,
xtick={20, 40, 60, 80, 100},
ytick={0.1, 0.2, 0.3, 0.4, 0.5},
xmajorgrids,
ymajorgrids,
legend style={nodes={scale=0.8}, legend cell align=left,at={(0.01,0.99)},anchor=north west, align=left, draw=white!15!black}
]
\addlegendimage{style={black,fill=blue,mark=none}}
\addlegendentry{$\alpha = 0.9$}
\addlegendimage{style={black,fill=red,mark=none}}
\addlegendentry{$\alpha = 0.7$}

\addplot[style={black,fill=blue,mark=none}] coordinates {(20,0.2899) (40,0.2957) (60, 0.3064) (80,0.3152) (100,0.3237)};

\addplot[style={black,fill=red,mark=none}] coordinates {(20,0.5118) (40,0.5163) (60,0.5253) (80,0.5354) (100,0.5499)};

\end{groupplot}

\end{tikzpicture}
	\caption{Spatial availability as a function of (a) $(\theta, \alpha)$ parameters (b) number of APs for $\theta = 0$ dB.}
	\label{fig:results_spatialanalysis} 
\end{figure}

In Section \ref{sec:spatialanalysis}, a framework for proposing $(\theta, \alpha, \Phi)$-available regions was introduced. As one would expect, the values selected for the mentioned parameters highly affect the achieved spatial availability, thus, to ensure an average insight over all possible AP locations, spatial averaging over a large number of deployments was conducted. In Fig. \ref{fig:results_spatialanalysis}, we highlight the effect of parameters $\theta$ and $\alpha$ along with the number of \ac{APs} on spatial availability. First, in Fig. (\ref{fig:results_spatialanalysis}a), the spatial availability $A_s$ of a randomly selected \ac{AP} is plotted as a function of $\theta$ for different confidence levels, $\alpha$. As expected, for increasing values of $\theta$ (or $\alpha$), the spatial availability of that \ac{AP} decreases, as the equivalent $(\theta, \alpha)$-available region reduces. 

Second, in Fig. (\ref{fig:results_spatialanalysis}b), $A_s$ is depicted as a function of the number of deployed \ac{APs} for two different confidence levels, when $\theta$=0 dB. The monotonically increasing fashion of $A_s$ as a function of $N$ for a given value of $\alpha$ is explained as follows: as the system becomes more densified with \ac{APs}, the Voronoi area of each \ac{AP} decreases due to the application of a distance-based association criterion and so does also its $(\theta, \alpha)$-available region, due to larger interference received. However, the latter region is less affected compared to the former, due to the stochastic nature of the region forming criterion together with the applied bisection-based approach for computing $(\theta, \alpha)$-available regions.
\subsection{Temporal Analysis}
\begin{figure} 
	\centering
	\input{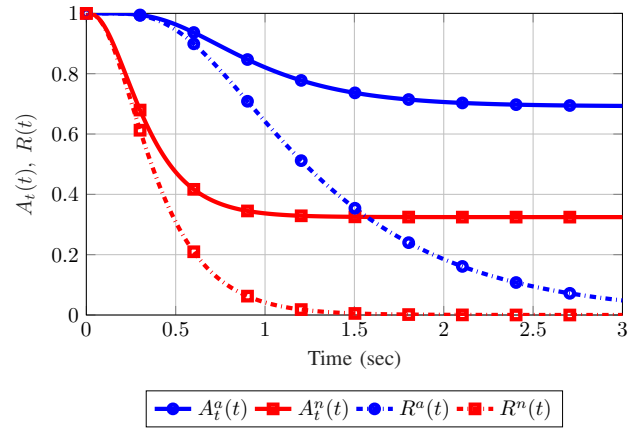}
	\caption{Temporal availability and reliability of $(\theta, \alpha)$-available and non-available regions for $A_s=0.7$.}
	\label{fig:results_temporalanalysis_1} 
\end{figure}

Based on the presented metrics in Section \ref{sec:temporalanalysis}, we investigate, in what follows, the temporal availability and reliability for the proposed access scheme. In Fig. \ref{fig:results_temporalanalysis_1}, the system's transient analysis is presented for $A_s=0.7$. Due to the spatially-dependent channel allocation proposed in eq. (\ref{eq:channel_access}), the time availability, $A_t^a(t)$ (reliability  $R^a(t)$) for a request originating from the $(\theta, \alpha)$-available region should be higher than the time availability $A_t^n(t)$ (reliability $R^n(t)$) of the  $(\theta, \alpha)$-non available region. This is explained due to the larger number of channels that can be utilized for the available region. It is noticeable that at $t=0$, all channels are available, thus, leading to time availability and reliability equal to one. Additionally, numerical results confirm that the time reliability is upper bounded by time availability, as well as that such a bound is time-dependent since it loosens over time till a maximum performance gap is reached which is then fixed over time.

\begin{figure} 
	\centering
	\input{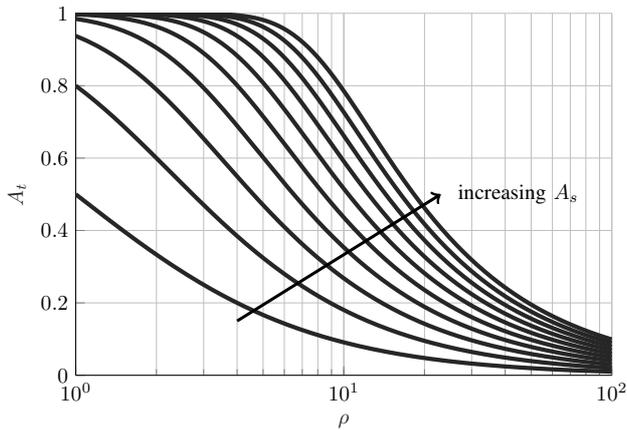}
	\caption{Steady state time availability, as a function of request arrivals to  service rate ratio, for increasing values of spatial availability ($A_s$) ranging from 0.1 to 1.}
	\label{fig:fig:results_temporalanalysis_2} 
\end{figure}

In Fig. (\ref{fig:fig:results_temporalanalysis_2}), the steady state analysis is illustrated for varying values of the arrival to service ratio $\rho$. As $\rho$ increases, the steady state temporal availability decreases; this occurs due to the fact that the available channels are less in such regimes. Also, as explained earlier, as a result of the	 adopted channel access scheme, larger values of spatial availability lead to higher time availability. It is, therefore, concluded that $\rho$ is a fundamental performance limitation factor, as for extremely large values of it, even a 100\% spatial availability is unable to be translated to high time availability.
\subsection{Joint Analysis}
\begin{figure} 
	\centering
	\begin{tikzpicture}[scale=0.8]

\begin{axis}[%
width=1*\columnwidth,
height=6cm,
scale only axis,
xmin=0,
xmax=1,
xlabel style={font=\color{white!15!black}},
xlabel={$A_s$},
ymin=0,
ymax=1,
ytick={0,0.2,0.4,0.6,0.8,1}, 
ylabel style={font=\color{white!15!black}},
ylabel={$A_t$},
axis background/.style={fill=white},
xmajorgrids,
ymajorgrids, legend cell align=left,
legend style={legend columns=-1, at={(1,0.98)},anchor=north east},
]
]
\addplot [color=blue, dashed, line width=2.0pt]
table[row sep=crcr]{%
0	0\\
0.1	0.111111111111111\\
0.2	0.219512195121951\\
0.3	0.324538258575198\\
0.4	0.425364758698092\\
0.5	0.520991696877558\\
0.6	0.610248114756048\\
0.7	0.691835300036913\\
0.8	0.764429738876318\\
0.9	0.826859172322988\\
1	0.878338935747049\\
};
\addlegendentry{$A^n_t$}
\addplot [color=blue, line width=2.0pt]
table[row sep=crcr]{%
0	0.878338935747049\\
0.1	0.826859172322988\\
0.2	0.764429738876318\\
0.3	0.691835300036913\\
0.4	0.610248114756048\\
0.5	0.520991696877558\\
0.6	0.425364758698092\\
0.7	0.324538258575198\\
0.8	0.219512195121951\\
0.9	0.111111111111111\\
1	0\\
};
\addlegendentry{$A^a_t$}
\addplot [color=red, dashed, line width=2.0pt]
table[row sep=crcr]{%
0	0\\
0.1	0.219512195121951\\
0.2	0.425364758698092\\
0.3	0.610248114756048\\
0.4	0.764429738876318\\
0.5	0.878338935747049\\
0.6	0.948593612287642\\
0.7	0.982779107833383\\
0.8	0.995470168283717\\
0.9	0.999055492450623\\
1	0.999841013561978\\
};
\addplot [color=red, line width=2.0pt]
table[row sep=crcr]{%
	0	0.999841013561978\\
	0.1	0.999055492450623\\
	0.2	0.995470168283717\\
	0.3	0.982779107833383\\
	0.4	0.948593612287642\\
	0.5	0.878338935747049\\
	0.6	0.764429738876318\\
	0.7	0.610248114756048\\
	0.8	0.425364758698092\\
	0.9	0.219512195121951\\
	1	0\\
};
\addplot [color=black, dashed, line width=2.0pt]
table[row sep=crcr]{%
0	0\\
0.1	0.324538258575198\\
0.2	0.610248114756048\\
0.3	0.826859172322988\\
0.4	0.948593612287642\\
0.5	0.990899111072123\\
0.6	0.999055492450623\\
0.7	0.999939437405946\\
0.8	0.999997446718659\\
0.9	0.999999925511194\\
1	0.99999999843439\\
};
\addplot [name path=A, color=black, line width=2.0pt]
table[row sep=crcr]{%
	0	0.99999999843439\\
	0.1	0.999999925511194\\
	0.2	0.999997446718659\\
	0.3	0.999939437405946\\
	0.4	0.999055492450623\\
	0.5	0.990899111072123\\
	0.6	0.948593612287642\\
	0.7	0.826859172322988\\
	0.8	0.610248114756048\\
	0.9	0.324538258575198\\
	1	0\\
};

\node[circle,fill=red,inner sep=2pt] at (axis cs:0.43123,0.8) {};
\node[circle,fill=red,inner sep=2pt] at (axis cs:0.56877,0.8) {};

\addplot[area legend, draw=red,draw opacity=0.01, fill=red, fill opacity=0.2, forget plot]
table[row sep=crcr] {%
x	y\\
0.43123	0\\
0.43123	0.8\\
0.56877	0.8\\
0.56877	0\\
}--cycle;

\node[circle,fill=black,inner sep=2pt] at (axis cs:0.28765,0.8) {};
\node[circle,fill=black,inner sep=2pt] at (axis cs:0.7124,0.8) {};
\addplot[area legend, draw=black, draw opacity=0.01, fill=black, fill opacity=0.2, forget plot]
table[row sep=crcr] {%
	x	y\\
	0.28765	0\\
	0.28765	0.8\\
	0.7124	0.8\\
	0.7124	0\\
}--cycle;

\draw[->, line width=0.5mm](axis cs:0.1,0.4) -- (axis cs:0.4,0.2);
\node[anchor=west] at (axis cs:0.4,0.18){decreasing $M$};

\end{axis}

\end{tikzpicture}%
	\caption{Steady state time availability as a function of spatial availability for varying numbers of channels ($M$ = (\textcolor{blue}{10}, \textcolor{red}{20}, 30)).}
	\label{fig:results_joint} 
\end{figure}
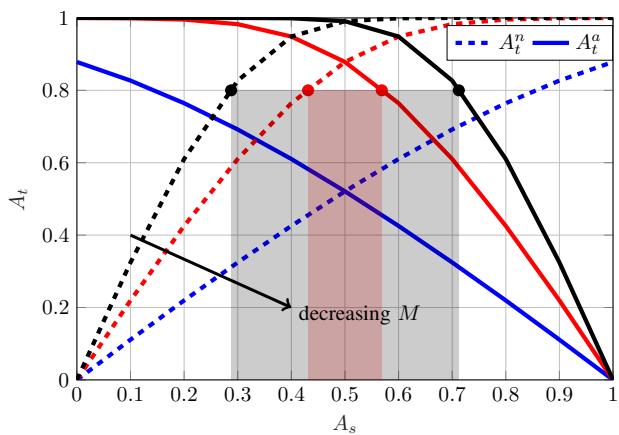

Finally, in Fig. (\ref{fig:results_joint}), the relationship between the steady-state time availability and the spatial availability is presented for different total channel numbers, $M$. First, we observe a symmetric time availability performance for a fixed $M$ between the $(\theta, \alpha)$-available and non-available regions, due to the proposed channel allocation scheme. For $As=0.5$, the number of channels allocated to each region will be the same, hence, leading to an identical time availability performance, as requests arrive uniformly in space. Additionally, fixing the value of $A_s$, time availability increases together with $M$. This result intuitively emphasizes the role of redundancy and provisioning in wireless systems. Equivalently, through our proposed space-time analysis, the minimum total number of channels needed to achieve a targeted temporal availability level can be identified. To further highlight this, a steady state time availability requirement of 0.8 is marked for $M=20$ and $M=30$ curves. As expected, the range of $A_s$ meeting the imposed requirement is larger in the latter case. This means that, a sufficient amount of resources can guarantee the time availability performance of multiple service classes.

\thispagestyle{empty}
\section{Conclusion}\label{sec:Conclusion}
In this paper we proposed a unified framework of computing temporal and spatial availability for a \ac{RAT}-agnostic system. A novel, service-relevant definition of spatial availability was introduced taking into account the probability to achieve a targeted \Ac{SIR} threshold with a given confidence level. Temporal availability was investigated considering a novel, space availability-driven channel access scheme based on the concept of channel provisioning, bringing up the coupled relation between spatial and temporal availability and reliability. The study is supported by numerical evaluation results which underline the impact of different system parameter values on space/time availability and time reliability, as well as the coupled nature of these metrics. Further research directions can be envisioned, focusing on the particularities of different RAN characteristics, the service traffic model, as well as the connectivity criteria and the allocation of other resources.
\section*{Acknowledgment}
The research leading to these results has been performed under the framework of the Horizon 2020 project ONE5G (ICT-760809) receiving funds from the European Union.
\thispagestyle{empty}
\bibliographystyle{./lib/IEEEtran.cls}
\bibliography{./literature/Literature_Local}

\thispagestyle{empty}

\end{document}